\documentclass[aps,twocolumn,pre,nofootinbib,groupaddress]{revtex4-1}
\usepackage{graphicx}
\newcommand{\MCtwo}{Microtechnology and Nanoscience, MC2,
Chalmers University of Technology, SE-412 96 G{\"o}teborg, Sweden}
\newcommand{\Stavanger}{Randaberg videreg{\aa}ende skole, Gr{\o}demveien 
70, NO-4070 Randaberg, Norway} 
\newcommand{\ontop}{{\mbox{\scriptsize top}}}
\newcommand{\hollow}{{\mbox{\scriptsize hollow}}}
\newcommand{\bridge}{{\mbox{\scriptsize bridge}}}

\begin{document}

\title{Methylbenzenes on graphene}

\author{{\O}yvind Borck}\affiliation{\Stavanger}
\author{Elsebeth Schr{\"o}der}\thanks{Corresponding author}\email{schroder@chalmers.se}%
\affiliation{\MCtwo}

\date{July 18, 2016}
\begin{abstract}
We present a theory study of the physisorption of the series of 
methylbenzenes (toluene, xylene and mesitylene), as well as benzene,
on graphene. This is relevant for the basic understanding of graphene used as
a material for sensors and as an idealized model for the carbon in active carbon filters.
The molecules are studied in a number of positions and 
orientations relative graphene, using density functional theory with the 
van der Waals functional vdW-DF. We focus on the vdW-DF1 and vdW-DF-cx 
functionals, and find that the binding energy of the molecules on 
graphene grows linearly with the number of methyl groups, at the rate of
0.09 eV per added methyl group.
\end{abstract}

\maketitle


\section{Introduction}

For environmental safety carbon-based filters play an important role 
for removal of toxic and hazardous substances, e.g., from the air
or drinking or waste water.
Smooth, defectless graphene may be used as a simplified model for the 
active carbon and similar material that is often used as filter material
in air and water filters.
Although defects and impurities play an important role in how active 
carbon acts as an adsorbent, already the calculated adsorption energy 
for adsorption on clean, perfect graphene will be an indication of the 
strength of the adsorption in the 
filters \cite{terranova12,SMLL:SMLL201500831,doi:10.1021/tx200339h,SMLL:SMLL201201125}.

At the same time, graphene-based sensors may be used for detection of specific 
molecules in gases and fluids. The sensor must respond selectively to the
various molecules. Both applications thus call for the need of understanding,
on a fundamental level, the interaction of a number of molecules with graphene. 

Here we focus on the adsorption of the group of methylbenzenes on graphene. 
We use of density functional theory (DFT) calculations
and the method vdW-DF \cite{dion04p246401,dion04p246401erratum,%
thonhauser07p125112,lee10p081101,berland14p035412,%
bearcoleluscthhy14,berland15p066501}
 to include the long-ranged dispersion interactions
that are crucial for physisorption. 
We find the adsorption energy and the structure (positions of atoms)
of the methylbenzenes when adsorbed in isolated positions on graphene.

With the same method, using functionals within the vdW-DF family, 
we have previously investigated the adsorption of other relatively 
small but important molecules on to graphene, such as benzene and 
naphthalene \cite{chakarova-kack06p146107}, 
phenol \cite{chakarova-kack06p155402}, 
adenine \cite{berland11p135001} and
the other nucleobases \cite{le12p424210}, 
chloroform and other trihalomethanes \cite{akesson12p174702}, 
methanol \cite{schroder13p871706},
and the first ten of the series of n-alkanes \cite{londero12p424212},
all at low coverage. 
These and similar results are useful as input for larger-scale 
force-field molecular dynamics calculations, as well as
providing fundamental knowledge on the binding properties at the 
single molecule level.

The outline for the rest of the paper is as follows:
Section II describes methylbenzenes, and 
Section III describes the method of computation, and 
the choices made in carrying out the calculations.
Section IV reports the results of our calculations,
Section V discusses our research results, and finally
we summarize the study in Section VI.

\section{Methylbenzenes}

\begin{figure}
\begin{center}
\includegraphics[width=0.06\textwidth]{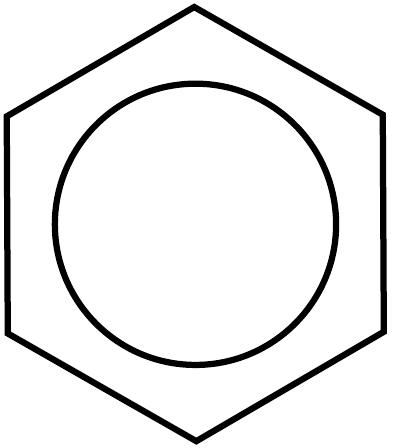}\hspace{0.03\textwidth}
\includegraphics[width=0.06\textwidth]{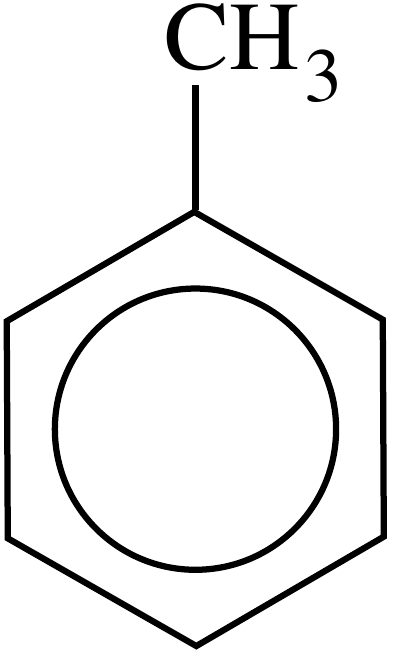}\hspace{0.03\textwidth}
\includegraphics[width=0.06\textwidth]{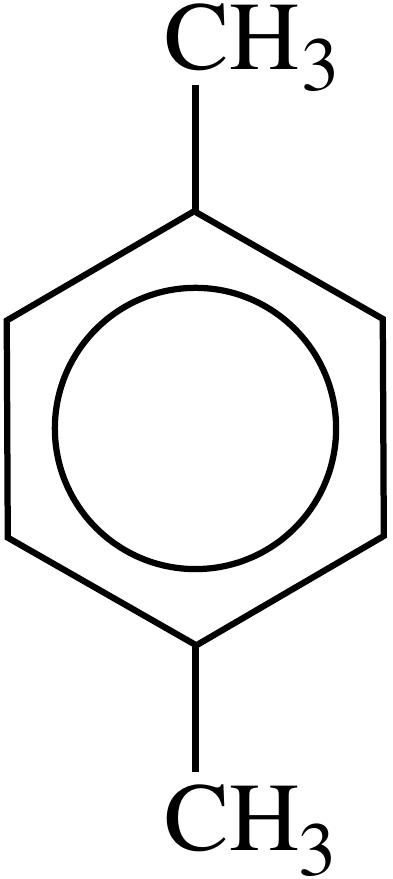}\hspace{0.03\textwidth}
\includegraphics[width=0.11\textwidth]{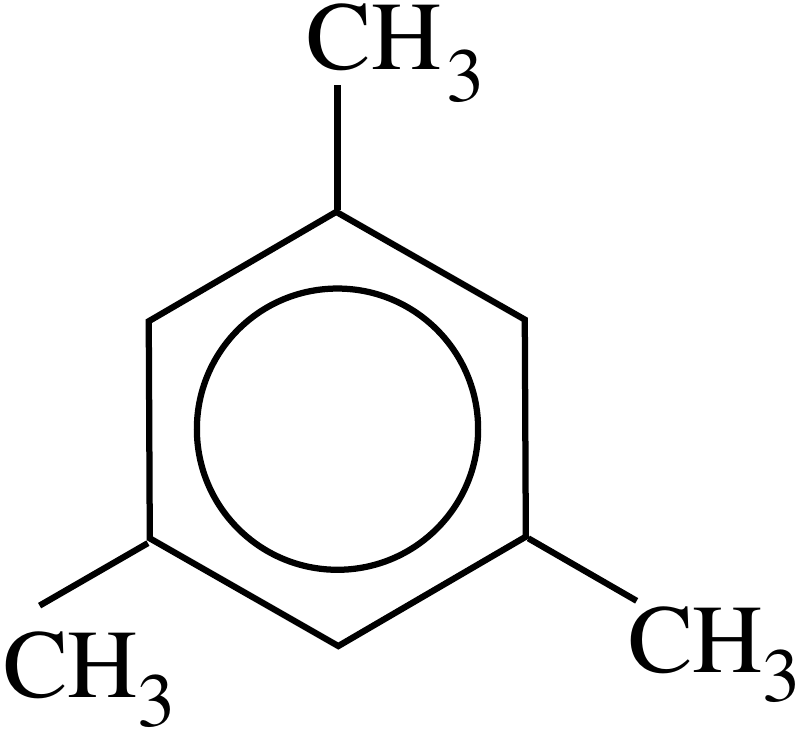}
\caption{\label{fig:btx}The atomic structures of benzene, and the methylbenzenes toluene, para-xylene,
and mesitylene.
}
\end{center}
\end{figure}

Methylbenzenes is a group of small, aromatic molecules that are volatile
and hazardous. 
They are benzene molecules that have one or more methyl groups attached.
We here focus on benzene and the methylbenzenes toluene,
para-xylene (1,4-dimethylbenzene)
and mesitylene (1,3,5-trimethylbenzene), with one, two, and three
methyl (CH$_3$) groups.
The atomic structures are shown in Figure \ref{fig:btx}.
To see the effects of isomers
we also study and compare  the adsorption energies of
xylene with the methyl groups
placed closer together: the ortho-xylene (1,2-dimethylbenzene)
and meta-xylene (1,3-dimethylbenzene).
Trimethylbenzene also exists as the isomers 1,2,3-trimethylbenzene 
(hemellitene) and 1,2,4-trimethylbenzene (pseudocumene), but these
will not be discussed here.
In this paper benzene is included as part of the group of 
methylbenzenes, although strictly taken benzene is not a methylbenzene.

Toluene is a colorless liquid, and is used as raw material for the 
industry and as a solvent. Toluene is
believed to be neurotoxic \cite{white97p1239}, and thus definitely not
suitable neither in drinking water nor should it be inhaled.
However, it is still much less toxic than benzene, and thus in some 
cases replaces benzene as an aromatic solvent. 
Xylene may also for some applications be used as a 
solvent even less toxic than toluene.

Graphene is a special material because it consists entirely of surface atoms.
The possible amount of adsorbed molecules per volume or weight of
carbon is therefore high, compared to other materials with similar
adsorption energies per adsorbed molecule.

\section{Method of computation}
The DFT calculations are carried out with the 
vdW-DF method \cite{dion04p246401,dion04p246401erratum,%
thonhauser07p125112,lee10p081101,berland15p066501} 
in which the exchange-correlation approximation
includes long-range dispersion interactions.
We mainly use the versions vdW-DF1 \cite{dion04p246401,dion04p246401erratum} 
and vdW-DF-cx \cite{berland14p035412},
although we also report some vdW-DF2 \cite{lee10p081101} results. 
The vdW-DF1 and vdW-DF2 calculations are 
carried out with the DFT code GPAW \cite{GPAWhttp,GPAW10} in the Atomic 
Simulation Environment (ASE) \cite{ASE02,ASEhttp}.
For calculations using the newer functional
vdW-DF-cx \cite{berland14p035412} 
we use the DFT code Quantum Espresso (QE) \cite{QEhttp,espresso} because 
vdW-DF-cx is not yet implemented in GPAW. All calculations use
a fast-Fourier-transform implementation of the central integral in
the nonlocal correlation calculations \cite{roso09}. 

\begin{figure}
\begin{center}
\includegraphics[width=0.25\textwidth]{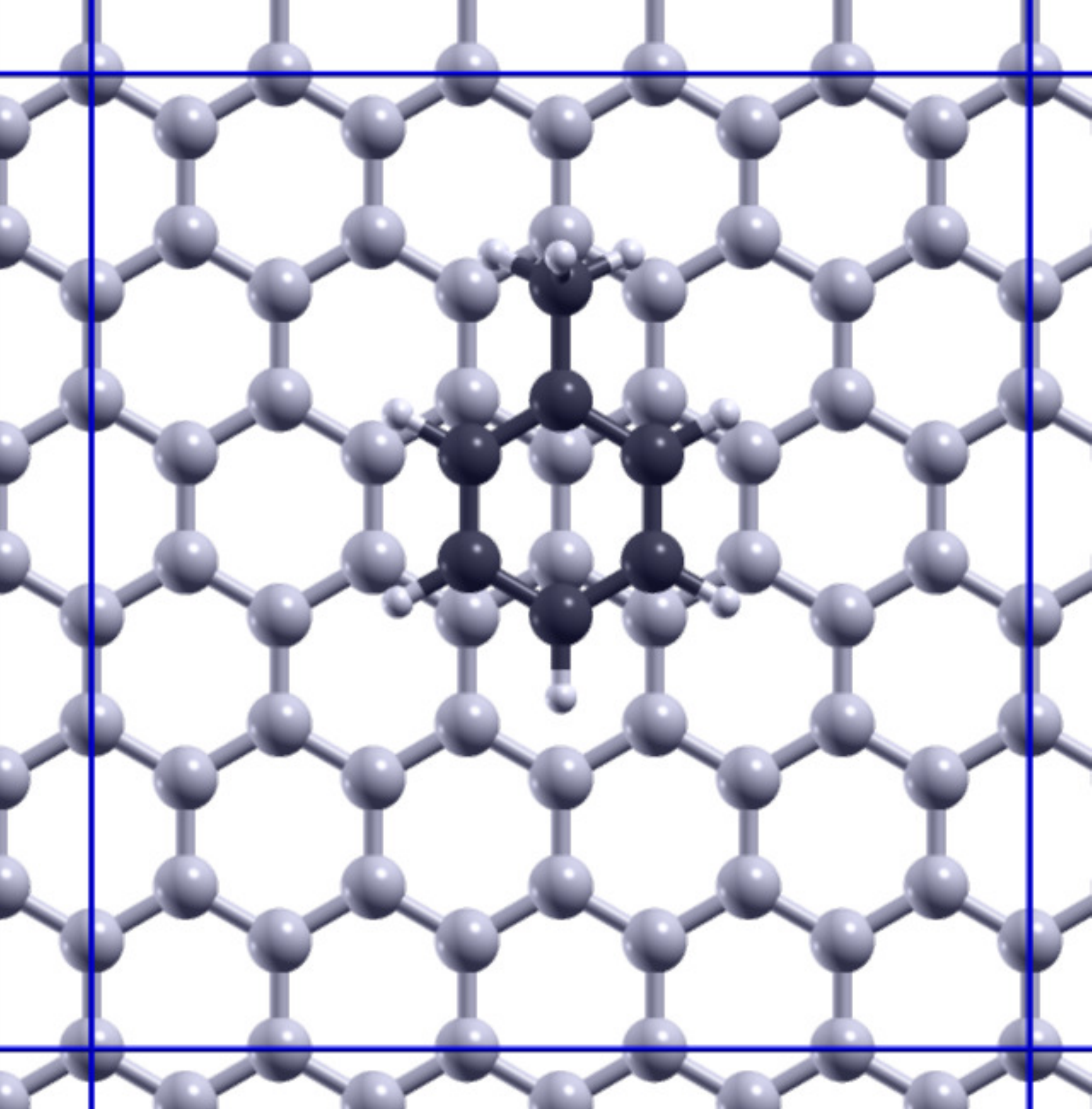}
\\[1 em]
\includegraphics[width=0.4\textwidth]{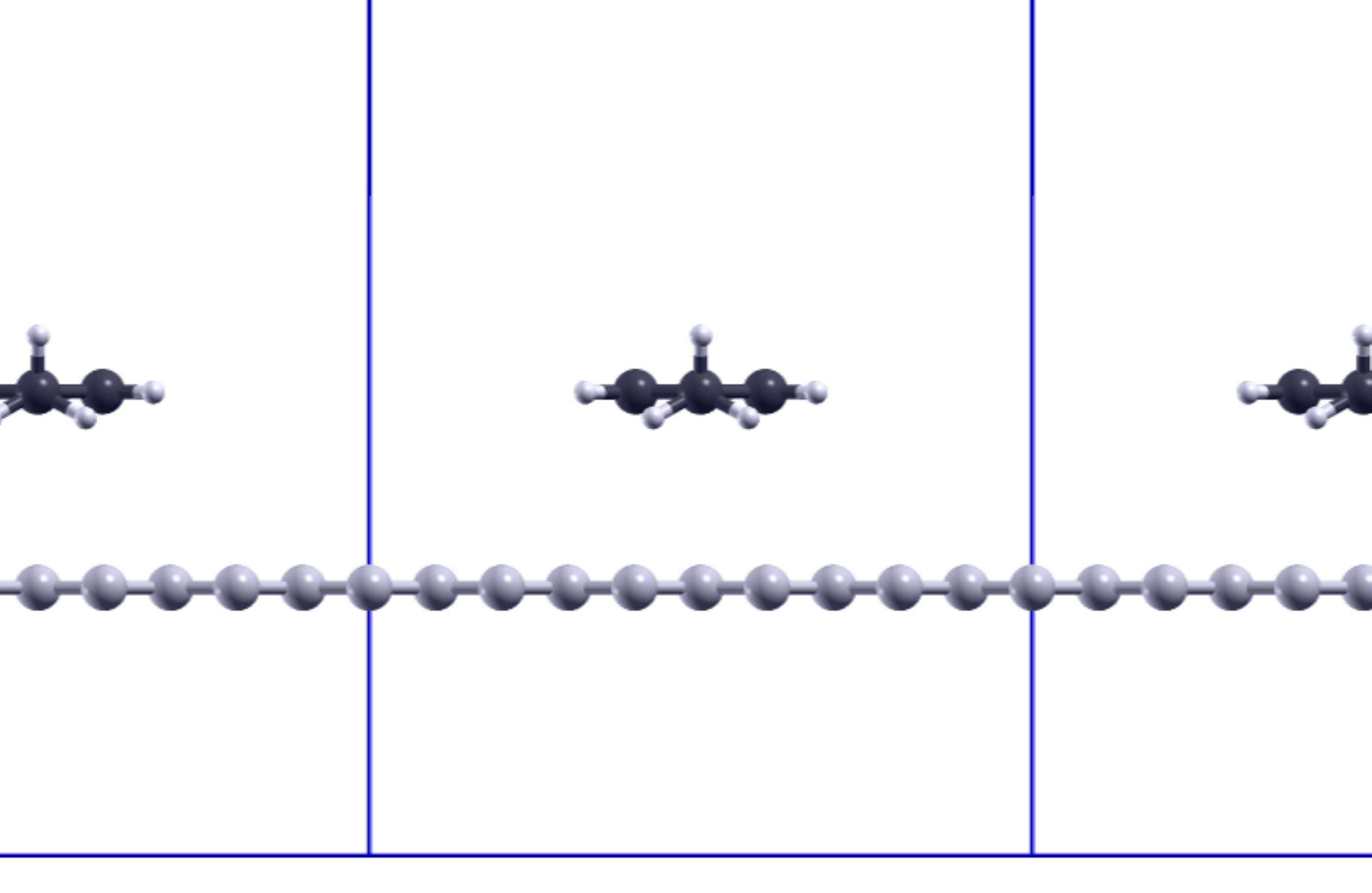}
\caption{\label{fig:slab}Sketch of unit cell for toluene adsorption.
The height of the unit cell is larger than shown in the lower panel. 
Light gray spheres 
are C atoms in graphene, dark gray spheres illustrate C atoms of
toluene, and small circles show positions of H on toluene. Shown is
the configuration with two H atoms of the methyl group H-tripod
pointing towards graphene (`edge') and with the center of the 
aromatic ring positioned on the bridge between two graphene C atoms 
(`bridge').
}
\end{center}
\end{figure}

In all calculations we use periodically repeated orthorhombic unit cells
of size $3\,\sqrt{3} a_g \times 5 a_g = 12.9$ {\AA}$\times 12.4$ {\AA} 
in the lateral plane, 
and 23 {\AA} in the direction perpendicular to the graphene plane, as 
illustrated in Figure \ref{fig:slab}. 
Here $a_g=\sqrt{3} \,a_c$, where $a_c=1.43$ {\AA} is the C-to-C distance  
 \cite{akesson12p174702} for the present calculations.
Each unit cell includes 60 C atoms in graphene and one methylbenzene molecule.
By the nature of the DFT calculations the system is in vacuum, and the 
calculations are carried out at zero temperature.

In the GPAW calculations we use a fast-Fourier transform grid with 
approximately 0.12 {\AA} between grid points for the wave functions, 
and half this distance for the electron density.
In Ref.~\cite{ericsson16preprint} we tested the convergence by using 
only 17\% of the grid points for toluene on graphene, along with the 
restriction to simple $\Gamma$-point sampling in $k$-space.
These rather significant changes to
the accuracy  resulted in a total of only 20 meV change
in adsorption energy, compared to the same calculation carried out
at the accuracy used in the present article \cite{ericsson16preprint}. 
The present grid point choices are therefore sufficient.   

The sampling of the wave functions and  electron density 
in the QE calculations is given by the energy 
cutoff values for the wave functions (electron density) 30 Ry (240 Ry),
yielding 0.22 {\AA} (0.08 {\AA}) between grid points for the wave 
function (electron density) sampling.

The atomic positions are determined by minimizing 
the forces acting on the atoms.
However, this is only a local optimization that cannot rotate the methyl groups or change 
the position of the aromatic ring on graphene, because the forces 
are too small, and we therefore use a series of different starting positions 
and orientation, as illustrated in the supplementary material.

\section{Results of the calculations}
Our main focus is the calculation of optimal adsorption configurations 
of the methylbenzenes, and their adsorption energies.
For each of the molecules we survey a number of 
systematically prepared configurations, all of which have their atomic 
positions further locally optimized, as explained in the previous section. 

We define the adsorption energy $E_a$ as the energy gained by moving 
the molecule from infinity to its adsorption position near graphene. 
Binding thus results in a positive value of $E_a$. 
In practice we use the distance 11.5 {\AA} from graphene as the position 
``far away''.
Our study provides adsorption energies, given in Table~\ref{tab:energies}
and the Supplementary material, as well as the changes in
adsorption energies when varying 
orientations, isomers, and positions relative to graphene.

\begin{table*}
\caption{\label{tab:energies}Adsorption energies $E_a$ for methylbenzenes
on graphene, using the functionals vdW-DF1 and vdW-DF2 in DFT program 
GPAW, and vdW-DF-cx in DFT program Quantum Espresso.
Also shown are experimental results from the literature.
For some combinations both the
configuration with all methyl groups having one H atom pointing towards
graphene (`corner') and configurations with all methyl groups having
two H atoms pointing towards graphene (`edge') are shown.
Numbers are in units of eV.
}

\begin{tabular}{lcccccc}
 \hline\hline
& \multicolumn{2}{c}{{vdW-DF1}}
& \multicolumn{1}{c}{{vdW-DF2}}
& \multicolumn{2}{c}{{vdW-DF-cx}}
& \multicolumn{1}{c}{{Experiments}}\\
                 &corner&edge&edge & corner & edge\\
  \hline
  benzene        &0.430&0.430&0.386&  0.511 & 0.511 & 0.44$^b$, $0.50\pm 0.08^c$\\ %
  toluene        &0.498$^a$&0.521&0.461&  0.599 & 0.620 & 0.52$^b$,
$0.71\pm 0.07^d$ \\
  para-xylene    &0.557&0.611&0.550&  0.666 & 0.721  \\
  meta-xylene    &     &0.612& \\
  ortho-xylene   &     &0.589& \\
  mesitylene    &0.602&0.701&0.632&  0.714 & 0.822  \\
\hline
$^a$Originally in Ref.~\cite{ericsson16preprint}.\\
$^b$Ref.~\cite{monkenbusch81p442}.\\
$^c$Ref.~\cite{ulbricht06p2931}.\\
$^d$Ref.~\cite{zacharia}.
\end{tabular}
\end{table*}

\begin{figure}
\begin{center}
\includegraphics[width=0.15\textwidth]{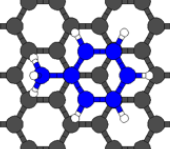}
\hspace{0.0\textwidth}
\includegraphics[width=0.15\textwidth]{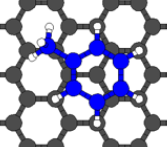}
\hspace{0.0\textwidth}
\includegraphics[width=0.15\textwidth]{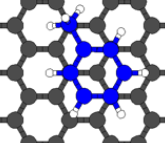}
\\[1 em]
\includegraphics[width=0.15\textwidth]{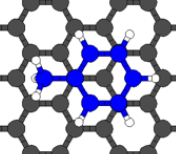}
\hspace{0.0\textwidth}
\includegraphics[width=0.15\textwidth]{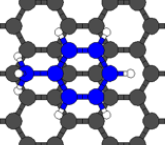}
\hspace{0.0\textwidth}
\includegraphics[width=0.15\textwidth]{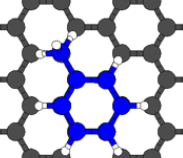}
\caption{\label{fig:positions}Examples of various positions of the
molecules on graphene, here
for toluene. Upper panels show `top' position of the aromatic ring,
with the methyl group turned (panels left to right) $0^\circ$, $30^\circ$, and $60^\circ$ around
the center of the aromatic ring. The lower panels
show `top', `bridge', and `hollow' positions of the aromatic ring.
The left lower panel shows the methyl group tripod turned so that 
a tripod `corner' is directed towards graphene, whereas the upper left 
panel shows the same configuration, except here a tripod `edge'
is directed towards graphene. 
All atomic positions have been locally optimized by minimizing
 the remaining Hellmann-Feynman forces on the atoms.
}
\end{center}
\end{figure}

All adsorption configurations considered, 
and their vdW-DF1 adsorption energies, are presented in 
the Supplementary material. Examples of configurations, for toluene, are
given in Figure \ref{fig:positions}. 
We study the effects of changing the details of the adsorption configurations
on the corresponding adsorption energies, and the trends 
in the adsorption energies  
for increasing number of methyl groups in the molecules. 
As discussed later, molecules like these have been found to
adsorb approximately flat on to graphene, when at low coverages \cite{grimaud2001}.

\subsection{Methyl group H-tripod rotations} 
Except benzene, all the molecules contain one or more methyl groups.
In each methyl group the triangle spanned by the three H atoms
can rotate around the axis connecting the methyl group C atom to 
the aromatic ring. In this work we consider two orientations of the tripod,
one where two of the H atoms are closer to graphene than the methyl group
C atom, which we term the `edge' configuration (Figure \ref{fig:slab}), 
and the `corner' orientation
where only one H atom is closer to graphene than the C atom.
The two orientations are also illustrated in the 
two left-most panels of Figure \ref{fig:positions}.

By comparing the adsorption energies 
when only the H tripod orientation differs in one or more methyl groups, 
we find that the `edge' configurations are energetically preferable.
We find a remarkably consistent cost for rotating a methyl
group from `edge' to `corner': 
Across eight such pairs of toluene configurations
we find that a tripod rotation from `edge' to `corner'
carries a cost of 22--28 meV, for para-xylene we find the cost 27 meV per
rotated H tripod, and the rotations in mesitylene come at a cost of
28--38 meV per tripod rotation. For mesitylene the cost per rotation
increases with the number of tripods already in `corner' orientation,
with 28 meV for the first rotation, 32 meV for the next, and 38 meV for
rotating also the third tripod from `edge' to `corner' orientation.

\subsection{Site of aromatic ring}
In graphite the formation energy depends on the stacking of the 
individual layers. Natural graphite is in AB stacking, which means 
that every other C atom in one layer is placed above a C atom in 
the neighboring layer, while the other half of the C atoms are above 
the middle of a ring.
It is thus natural, for flat aromatic molecules like the methylbenzenes,
to expect positions that involve placing the aromatic ring 
either above an atom in graphene (`top' position)
 or above the middle of a ring in graphene (`hollow' position). 
For benzene it is known that the `top' position is preferable
\cite{terranova12,chakarova05p054102,chakarova-kack06p146107,chakarova-kack10p013017}.
In addition, we also include calculations with the aromatic ring 
centered on the bridge between two neighboring
graphene C atoms (`bridge' position).
All positions are illustrated in Figure \ref{fig:positions}, lower panels. 

In most of our calculations we find `top' sites to be preferable,
with the `bridge' sites close in energy.
Distinguishing configurations with H tripod orientations
`edge' or `corner' we find that for toluene, going from the 
`top' to the `hollow' configuration costs 17--21 meV in 
adsorption energy, see Table~\ref{tab:diffenergies}.
The site-change cost is highest for the `corner' orientation of the tripod.
Changing
instead from `top' to the energetically intermediate `bridge' site
yields a vanishing (less than 1 meV) cost.  
Here, we have chosen the energetically best configurations for the 
`top', `bridge', and `hollow' adsorption configuration 
(see Supplementary information for 
numbers) in the comparison,
see Table~\ref{tab:diffenergies}. 

For the molecules with more than one methyl group the `bridge' site
tends to also be almost as good as the `top' position, with insignificant
energy differences (Table \ref{tab:diffenergies}).
The difference in adsorption energy between `top' and `hollow' positions
is for all the xylenes and for mesitylene around 24 meV or less.

\begin{table}
\caption{\label{tab:diffenergies}Difference in adsorption
energies $E_a$ for methylbenzenes
on graphene calculated with vdW-DF1.
In all cases we use the best adsorption energy that has the indicated
adsorption site and the indicated orientation of all
methyl group H tripods. Numbers are in units of meV.
All energy differences less than approximately 10 meV are insignificant.
}

\begin{tabular}{lcccc}
 \hline\hline
& \multicolumn{2}{c}{{$E_a^\ontop-E_a^\hollow$}}
& \multicolumn{2}{c}{{$E_a^\ontop-E_a^\bridge$}}
\\
                 &corner&edge&corner& edge\\
  \hline
  toluene        &   21&   17&    2&    1 \\
  para-xylene    &     &   21&     & $<1$ \\
  meta-xylene    &     &   23&     &    3\\
  ortho-xylene   &     &   23&     &    7\\
  mesitylene    &     &   24&     &    3 \\
\hline
\end{tabular}
\end{table}

\subsection{Rotations around the aromatic center}
In each adsorption site, the molecule may be rotated around the center 
of the aromatic ring. For toluene in the `top' site there are three 
orientations of the methyl group that carries some symmetry. They are 
obtained by a $30^\circ$ and $60^\circ$ rotation around the 
aromatic center, as illustrated in the top panels of 
Figure \ref{fig:positions}. For the `hollow' site the $60^\circ$ rotation
is equivalent to the unrotated orientation and thus only two
different orientations (none and $30^\circ$ rotation) are considered.
For the `bridge' site we study both the $30^\circ$ and $60^\circ$ 
rotated configurations, even though they carry less symmetry than the 
unrotated configuration, as shown in the middle lower panel of
Figure \ref{fig:positions}. All of these eight configurations for 
sites `top', `hollow', and `bridge' have 
two versions, with the `edge' and the `corner' orientation of the
methyl group H tripod. We find that in all cases, rotation of the molecule 
around the aromatic center
(keeping the H tripod the same) results
in energy changes of less than 8 meV, and in most cases even less. 

For the xylenes the effect of rotation around the aromatic center
is in all cases 7 meV or less. 
For mesitylene there is no change 
in energy ($<1$ meV) in any of the chosen rotations. Thus, with
the effect of rotation being 8 meV or less for all methylbenzenes we
find that it is reasonable to 
ignore the effect of rotations of the molecules.

\subsection{Optimal adsorption energies}
Focusing now on the optimal adsorption energy, irrespectable of
adsorption site and rotation of the molecule around the aromatic 
center, we
find the adsorption energies listed in Table \ref{tab:energies}. 

\begin{figure}
\begin{center}
\includegraphics[width=0.45\textwidth]{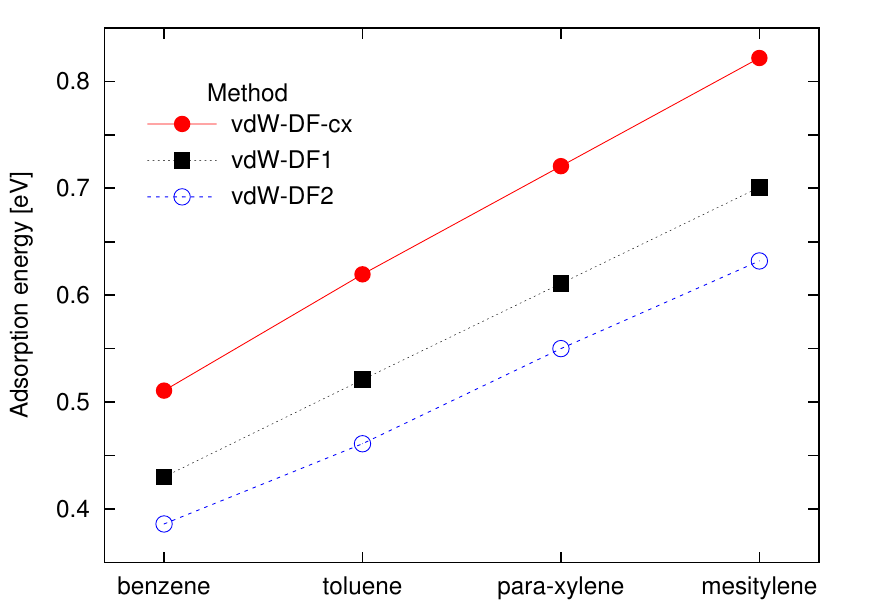}
\caption{\label{fig:btxenergies}Adsorption energy of benzene, 
and the methylbenzenes toluene, para-xylene,
and mesitylene, calculated using the vdW-DF method in versions vdW-DF-cx, 
vdW-DF1, and vdW-DF2. All results are for the optimal configuration,
and for xylene and tri-methylbenzene we choose the structural isomer that
has its
methyl groups evenly distributed around the aromatic ring (para-xylene and
mesitylene).
}
\end{center}
\end{figure}

From Table \ref{tab:energies} we find that as
 the number of methyl groups in the  molecule grows, 
from benzene to mesitylene, the adsorption energy per
molecule also grows. This is shown in Figure \ref{fig:btxenergies} for 
the `edge' configuration. 
For dimethylbenzene and trimethylbenzene we use the isomers
with evenly distributed methyl groups (para-xylene and mesitylene).
In Figure \ref{fig:btxenergies} we plot, for each molecule and version 
of vdW-DF, the highest adsorption energy of all the calculated 
positions and orientations. 
The distance of the molecule
aromatic center to the graphene plane is similar for all the molecules:
approximately 3.6 {\AA} for vdW-DF1, 3.5--3.6 {\AA} for vdW-DF2, and 
3.3--3.4 
{\AA} for vdW-DF-cx, all for the `edge' configuration. Distances are
slightly larger
for the `corner' configuration. The toluene molecule is adsorbed
with the aromatic ring with a small angle (a slight tilt) to the plane of
graphene, due to the asymmetry of the toluene molecule.

\section{Discussion of results}

In the previous section we present adsorption energies 
for the group of methylbenzene molecules, in several adsorption 
configurations. 
We also extract the effects on the adsorption energies when changing 
the adsorption geometry in various ways, such as rotating the molecule 
around its aromatic center, rotating the H tripod of the methyl group(s) 
and translating the molecule along graphene.
The vdW-DF1 is constructed with exchange from the revPBE functional \cite{zhang98p890}.
Due to the overly repulsive revPBE exchange, the
vdW-DF1 is prone
to overestimating the adsorption distances and underestimating  
binding energies.
In vdW-DF2 the inner functional \cite{hybesc14} is replaced
with the goal of working better for small molecules, like 
the G22 set of molecules \cite{curtiss91p7221} and other systems with 
atoms or small molecules. It has been shown
to work well for such systems 
\cite{lee10p081101,lee10p155461,lee11p193408,chen12p424211}.
However, part of the system considered here is an extended system 
(graphene) which was not a target system for the development of vdW-DF2.
In vdW-DF-cx the exchange functional is chosen so as to match the 
inner functional \cite{berland14p035412,hybesc14} and to observe important
physics rules. In general, vdW-DF-cx
improves the adsorption distances and energies compared to vdW-DF1.
As previously seen in adsorption systems \cite{bearcoleluscthhy14,thonhauser15p136402},
the vdW-DF-cx 
provides larger overall binding energies than for vdW-DF1, here with energies
0.08--0.12 eV larger per molecule (Table \ref{tab:energies}).

The adsorption energies $E_a$ of Table \ref{tab:energies} are plotted 
in Figure \ref{fig:btxenergies} as a function of growing number of 
methyl groups, for the `edge' configuration of the
H tripods. We see that the increase in $E_a$ is almost
linear with number of methyl groups, with an offset, corresponding to 
the adsorption energy of benzene, Table \ref{tab:energies}.
For each methyl group added 
an H atom is also removed, thus the net gain of atoms is one C  and 
two H atoms.
One of us found in an earlier study \cite{londero12p424212}, by use of 
vdW-DF1, that in adsorption of the linear n-alkanes on graphene 
the gain in adsorption energy per additional CH$_2$ group added to the length 
of the alkane molecule is 0.075 eV. 
This fits reasonably well with the increase found here when we use vdW-DF1: 
$E_a$ increases by 0.090 eV per added methyl group. 
For the vdW-DF-cx calculations the corresponding increase is also approximately
0.09 eV per methyl group, while the vdW-DF2 results vary more and are
generally
0.04--0.06 eV smaller than the vdW-DF1 results.
The rather systematic increase per added methyl group for vdW-DF-cx and vdW-DF1
suggests that in modeling systems with large or many molecules,
the methyl group may be modeled as an entity.

The unit cell in the calculations contains one molecule, in addition
to graphene. However, because of the periodic boundary conditions
each molecule has neighbors at approximately 12 {\AA} distance
(the length of the sides of the unit cell).
This is far from full coverage. In Ref.~\cite{monkenbusch81p442} the area occupied by an
adsorbed toluene molecule on graphite is measured to be 46 {\AA}$^2$.
In our orthorhombic unit cell of lateral size 12.9 {\AA} $\times$ 12.4 {\AA} area
one molecule per unit cell thus leads to 
an estimated coverage $0.29$ ML. 
In Ref.~\cite{klomkliang12p5320}, on the other hand, Monte Carlo simulations are used to find
the coverage concentration 4.30 $\mu$mol/m$^2$, which corresponds to an
area per molecule 38.6 {\AA}$^2$. From this number, we can estimate our coverage
to approximately 0.24 ML, in any case far from full coverage. 

At sparse coverages, such as here, it is favorable for the relatively flat methylbenzenes
to orient the aromatic ring
approximately parallel to graphene, in order to maximize the area of interaction.
On the other hand, at high coverages, the total interaction with graphene and with 
neighboring molecules is maximized if the molecules tilt 
or stand up perpendicular to graphene 
This was shown for close to full coverage in empirical Monte Carlo
simulations of benzene, toluene, and para-xylene on graphene \cite{klomkliang12p5320},
and in experiments for other flat molecules in Ref.~\cite{grimaud2001}.
At a coverage of approximately 0.24--0.29 ML the molecules in our calculations 
are expected to benefit energetically from being in a position parallel to graphene,
which is the orientation we consider here.

Optimizing how the methylbenzenes
are positioned relative to graphene there are several 
minima in the binding energy.
As described in the previous section, the (center of) the aromatic 
ring can be positioned on top of 
a graphene C atom (`top'), on top of an aromatic ring in graphene
(`hollow'), on top
of the bridge between two nearest-neighbor graphene C atoms (`bridge'), or any
position in between these. The ring can be rotated around its center
and thus change the position of the methyl-group C atom(s) relative 
to graphene. Further, the methyl group H-atom tripod can be rotated around
its C atom such as to have one or two H-atoms pointing towards graphene.
Several of these changes in positions are surveyed in the Supplemental
material, here we extract and discuss their energy differences.
 
As explained in the results section, the `edge' orientation of the methyl
group is preferable, gaining 22--38 meV per methyl group compared to the 
`corner' orientation. This is natural, as by having the H-tripod 
edges pointing to graphene the whole molecule can move closer to 
graphene, and it thus obtains a slightly larger adsorption energy.

Comparing the position of the aromatic ring relative to graphene, 
with everything else kept the same, we find that the `top' or `bridge' 
position is most favorable. Changing the ring position to the `hollow' 
position comes at a cost of about 20 meV.
As found for chloroform on graphene \cite{akesson12p174702}, the energy difference
for the various adsorption sites is small, and the corrugation of 
graphene also for the methylbenzenes studied here is so small that 
it only takes little kinetic energy to overcome the energy barriers 
for moving along graphene once the molecule is adsorbed on graphene.
At room temperature the methylbenzenes are free to move, both 
translational and rotational.

We previously reported the results of toluene on graphene, by use of vdW-DF1,
obtained by visiting high school students who worked in a research project
in our group \cite{ericsson16preprint}.
For time reasons the students used a lower quality on the convergence parameters,
they used the `corner' orientation of the methyl H-tripod,
and they did not explore the energy landscape of various positions and orientations
of toluene on graphene. However, toluene was (as here) placed with the aromatic
ring parallel to graphene and atomic positions were allowed to relax locally.
The students obtained the adsorption energy 0.479 eV per molecule,
whereas the present study gives adsorption energies for toluene 
in orientation `corner' in the range
0.476--0.498 eV, using vdW-DF1 for their calculations
(Table~\ref{tab:energies}).
Thus, the lower accuracy of the students' calculations
only affected the binding energy by 0.01--0.02 eV (1--2 kJ/mol), or less than 4\%,
compared to higher accuracy in the present study.

In an early stage of the vdW-DF development one of us calculated the 
adsorption energy of benzene on graphene, using vdW-DF1 but without
local optimization of atomic positions. The DFT code used was dacapo 
in a locally supplied non-selfconsistent version for the dispersion interaction
in vdW-DF1, used after a consistent GGA calculation of the electron density. 
Those calculations showed an adsorption energy for benzene 
0.495 eV (0.763 eV for naphthalene) \cite{chakarova-kack06p146107}.

For toluene and benzene experimental measurements 
of the heat of adsorption are available from Refs.~\cite{zacharia,ulbricht06p2931,monkenbusch81p442},
included in Table \ref{tab:energies}. 
The values indicate that the calculated results are reasonable,
taking into account that the experiments are for systems that
are not in vacuum, are at non-zero temperature, and measured at
approximately one monolayer (ML) coverage, much denser than the
present calculations. In Ref.~\cite{ericsson16preprint} we estimated the zero-point
motion of toluene to be just a few meV, which we can therefore ignore here. 

In the calculations presented here the adsorption systems
are in vacuum, which is not the situation in water or
air filters.
One of us has earlier studied how the presence of water molecules
affects the adsorption results of, e.g., chloroform on graphene
or graphene oxide \cite{akesson12p174702,kuisma16p}.
As argued in Ref.~\cite{kuisma16p},  if the adsorption
energy is affected by water, that effect can only show
in the part of the energy containing the nonlocal part of the
correlation interaction. On the other hand, that nonlocal part of the energy is 
found as a sum of poles in a contour integral in the space of complex frequencies,
the sum starting at frequencies much higher than the range where the index of 
refraction for water differs from unity. The water molecules
do not engage in the vibrations and thus do not change the nonlocal interaction.

\section{Summary}
We report on a density functional study of methylbenzenes adsorbed on
graphene. We find that although some adsorption configurations are 
energetically better than others, the energy differences are all small.
Changing  the orientation of the H tripod on the methyl group(s) of the
molecules gives rise to the largest energy difference, which is 
approximately in the range 22--38 meV per methyl group, 
followed by the positioning of the
center of the aromatic ring on graphene, showing a difference of only 20 meV,
which means that at room temperature this difference can be ignored. 
Rotations around the center of the aromatic ring are so small (8 meV or less)
that they may be ignored entirely, even at low temperature. 
Including also benzene in the study we show that the adsorption energy
per molecule increases approximately linearly with number of methyl groups in the
molecule. We use the vdW-DF method, mainly in the original version vdW-DF1,
but we also use the vdW-DF-cx and vdW-DF2 versions. As expected, the adsorption energies
differ between these, with the most reliable method vdW-DF-cx yielding the
largest adsorption energies. 

\acknowledgments
Support from the Swedish Research Council (VR) is gratefully 
acknowledged.
The computations were in part performed on resources 
at Chalmers Centre for Computational Science and Engineering (C3SE) 
provided by the Swedish National Infrastructure for Computing (SNIC).


\begin{thebibliography}{42}%
\makeatletter
\providecommand \@ifxundefined [1]{%
 \@ifx{#1\undefined}
}%
\providecommand \@ifnum [1]{%
 \ifnum #1\expandafter \@firstoftwo
 \else \expandafter \@secondoftwo
 \fi
}%
\providecommand \@ifx [1]{%
 \ifx #1\expandafter \@firstoftwo
 \else \expandafter \@secondoftwo
 \fi
}%
\providecommand \natexlab [1]{#1}%
\providecommand \enquote  [1]{``#1''}%
\providecommand \bibnamefont  [1]{#1}%
\providecommand \bibfnamefont [1]{#1}%
\providecommand \citenamefont [1]{#1}%
\providecommand \href@noop [0]{\@secondoftwo}%
\providecommand \href [0]{\begingroup \@sanitize@url \@href}%
\providecommand \@href[1]{\@@startlink{#1}\@@href}%
\providecommand \@@href[1]{\endgroup#1\@@endlink}%
\providecommand \@sanitize@url [0]{\catcode `\\12\catcode `\$12\catcode
  `\&12\catcode `\#12\catcode `\^12\catcode `\_12\catcode `\%12\relax}%
\providecommand \@@startlink[1]{}%
\providecommand \@@endlink[0]{}%
\providecommand \url  [0]{\begingroup\@sanitize@url \@url }%
\providecommand \@url [1]{\endgroup\@href {#1}{\urlprefix }}%
\providecommand \urlprefix  [0]{URL }%
\providecommand \Eprint [0]{\href }%
\providecommand \doibase [0]{http://dx.doi.org/}%
\providecommand \selectlanguage [0]{\@gobble}%
\providecommand \bibinfo  [0]{\@secondoftwo}%
\providecommand \bibfield  [0]{\@secondoftwo}%
\providecommand \translation [1]{[#1]}%
\providecommand \BibitemOpen [0]{}%
\providecommand \bibitemStop [0]{}%
\providecommand \bibitemNoStop [0]{.\EOS\space}%
\providecommand \EOS [0]{\spacefactor3000\relax}%
\providecommand \BibitemShut  [1]{\csname bibitem#1\endcsname}%
\let\auto@bib@innerbib\@empty
\bibitem [{\citenamefont {Terranova}\ \emph {et~al.}(2012)\citenamefont
  {Terranova}, \citenamefont {Orlanducci},\ and\ \citenamefont
  {Rossi}}]{terranova12}%
  \BibitemOpen
  \bibinfo {editor} {\bibfnamefont {M.~L.}\ \bibnamefont {Terranova}}, \bibinfo
  {editor} {\bibfnamefont {S.}~\bibnamefont {Orlanducci}}, \ and\ \bibinfo
  {editor} {\bibfnamefont {M.}~\bibnamefont {Rossi}},\ eds.,\ \href@noop {}
  {\emph {\bibinfo {title} {Carbon Nanomaterials for Gas Adsorption}}}\
  (\bibinfo  {publisher} {CRC Press},\ \bibinfo {year} {2012})\BibitemShut
  {NoStop}%
\bibitem [{\citenamefont {Mao}\ \emph {et~al.}(2015)\citenamefont {Mao},
  \citenamefont {Chang}, \citenamefont {Zhou},\ and\ \citenamefont
  {Chen}}]{SMLL:SMLL201500831}%
  \BibitemOpen
  \bibfield  {author} {\bibinfo {author} {\bibfnamefont {S.}~\bibnamefont
  {Mao}}, \bibinfo {author} {\bibfnamefont {J.}~\bibnamefont {Chang}}, \bibinfo
  {author} {\bibfnamefont {G.}~\bibnamefont {Zhou}}, \ and\ \bibinfo {author}
  {\bibfnamefont {J.}~\bibnamefont {Chen}},\ }\href {\doibase
  10.1002/smll.201500831} {\bibfield  {journal} {\bibinfo  {journal} {Small}\
  }\textbf {\bibinfo {volume} {11}},\ \bibinfo {pages} {5336} (\bibinfo {year}
  {2015})}\BibitemShut {NoStop}%
\bibitem [{\citenamefont {Sanchez}\ \emph {et~al.}(2012)\citenamefont
  {Sanchez}, \citenamefont {Jachak}, \citenamefont {Hurt},\ and\ \citenamefont
  {Kane}}]{doi:10.1021/tx200339h}%
  \BibitemOpen
  \bibfield  {author} {\bibinfo {author} {\bibfnamefont {V.~C.}\ \bibnamefont
  {Sanchez}}, \bibinfo {author} {\bibfnamefont {A.}~\bibnamefont {Jachak}},
  \bibinfo {author} {\bibfnamefont {R.~H.}\ \bibnamefont {Hurt}}, \ and\
  \bibinfo {author} {\bibfnamefont {A.~B.}\ \bibnamefont {Kane}},\ }\href
  {\doibase 10.1021/tx200339h} {\bibfield  {journal} {\bibinfo  {journal}
  {Chem. Res. Toxicol.}\ }\textbf {\bibinfo {volume} {25}},\ \bibinfo {pages}
  {15} (\bibinfo {year} {2012})}\BibitemShut {NoStop}%
\bibitem [{\citenamefont {Maliyekkal}\ \emph {et~al.}(2013)\citenamefont
  {Maliyekkal}, \citenamefont {Sreeprasad}, \citenamefont {Krishnan},
  \citenamefont {Kouser}, \citenamefont {Mishra}, \citenamefont {Waghmare},\
  and\ \citenamefont {Pradeep}}]{SMLL:SMLL201201125}%
  \BibitemOpen
  \bibfield  {author} {\bibinfo {author} {\bibfnamefont {S.~M.}\ \bibnamefont
  {Maliyekkal}}, \bibinfo {author} {\bibfnamefont {T.~S.}\ \bibnamefont
  {Sreeprasad}}, \bibinfo {author} {\bibfnamefont {D.}~\bibnamefont
  {Krishnan}}, \bibinfo {author} {\bibfnamefont {S.}~\bibnamefont {Kouser}},
  \bibinfo {author} {\bibfnamefont {A.~K.}\ \bibnamefont {Mishra}}, \bibinfo
  {author} {\bibfnamefont {U.~V.}\ \bibnamefont {Waghmare}}, \ and\ \bibinfo
  {author} {\bibfnamefont {T.}~\bibnamefont {Pradeep}},\ }\href {\doibase
  10.1002/smll.201201125} {\bibfield  {journal} {\bibinfo  {journal} {Small}\
  }\textbf {\bibinfo {volume} {9}},\ \bibinfo {pages} {273} (\bibinfo {year}
  {2013})}\BibitemShut {NoStop}%
\bibitem [{\citenamefont {Dion}\ \emph {et~al.}(2004)\citenamefont {Dion},
  \citenamefont {Rydberg}, \citenamefont {Schr{\"o}der}, \citenamefont
  {Langreth},\ and\ \citenamefont {Lundqvist}}]{dion04p246401}%
  \BibitemOpen
  \bibfield  {author} {\bibinfo {author} {\bibfnamefont {M.}~\bibnamefont
  {Dion}}, \bibinfo {author} {\bibfnamefont {H.}~\bibnamefont {Rydberg}},
  \bibinfo {author} {\bibfnamefont {E.}~\bibnamefont {Schr{\"o}der}}, \bibinfo
  {author} {\bibfnamefont {D.~C.}\ \bibnamefont {Langreth}}, \ and\ \bibinfo
  {author} {\bibfnamefont {B.~I.}\ \bibnamefont {Lundqvist}},\ }\href {\doibase
  10.1103/PhysRevLett.92.246401} {\bibfield  {journal} {\bibinfo  {journal}
  {Phys. Rev. Lett.}\ }\textbf {\bibinfo {volume} {92}},\ \bibinfo {pages}
  {246401} (\bibinfo {year} {2004})}\BibitemShut {NoStop}%
\bibitem [{\citenamefont {Dion}\ \emph {et~al.}(2005)\citenamefont {Dion},
  \citenamefont {Rydberg}, \citenamefont {Schr{\"o}der}, \citenamefont
  {Langreth},\ and\ \citenamefont {Lundqvist}}]{dion04p246401erratum}%
  \BibitemOpen
  \bibfield  {author} {\bibinfo {author} {\bibfnamefont {M.}~\bibnamefont
  {Dion}}, \bibinfo {author} {\bibfnamefont {H.}~\bibnamefont {Rydberg}},
  \bibinfo {author} {\bibfnamefont {E.}~\bibnamefont {Schr{\"o}der}}, \bibinfo
  {author} {\bibfnamefont {D.~C.}\ \bibnamefont {Langreth}}, \ and\ \bibinfo
  {author} {\bibfnamefont {B.~I.}\ \bibnamefont {Lundqvist}},\ }\href {\doibase
  10.1103/PhysRevLett.95.109902} {\bibfield  {journal} {\bibinfo  {journal}
  {Phys. Rev. Lett.}\ }\textbf {\bibinfo {volume} {95}},\ \bibinfo {pages}
  {109902} (\bibinfo {year} {2005})}\BibitemShut {NoStop}%
\bibitem [{\citenamefont {Thonhauser}\ \emph {et~al.}(2007)\citenamefont
  {Thonhauser}, \citenamefont {Cooper}, \citenamefont {Li}, \citenamefont
  {Puzder}, \citenamefont {Hyldgaard},\ and\ \citenamefont
  {Langreth}}]{thonhauser07p125112}%
  \BibitemOpen
  \bibfield  {author} {\bibinfo {author} {\bibfnamefont {T.}~\bibnamefont
  {Thonhauser}}, \bibinfo {author} {\bibfnamefont {V.~R.}\ \bibnamefont
  {Cooper}}, \bibinfo {author} {\bibfnamefont {S.}~\bibnamefont {Li}}, \bibinfo
  {author} {\bibfnamefont {A.}~\bibnamefont {Puzder}}, \bibinfo {author}
  {\bibfnamefont {P.}~\bibnamefont {Hyldgaard}}, \ and\ \bibinfo {author}
  {\bibfnamefont {D.~C.}\ \bibnamefont {Langreth}},\ }\href {\doibase
  10.1103/PhysRevB.76.125112} {\bibfield  {journal} {\bibinfo  {journal} {Phys.
  Rev. B.}\ }\textbf {\bibinfo {volume} {76}},\ \bibinfo {pages} {125112}
  (\bibinfo {year} {2007})}\BibitemShut {NoStop}%
\bibitem [{\citenamefont {Lee}\ \emph {et~al.}(2010{\natexlab{a}})\citenamefont
  {Lee}, \citenamefont {Murray}, \citenamefont {Kong}, \citenamefont
  {Lundqvist},\ and\ \citenamefont {Langreth}}]{lee10p081101}%
  \BibitemOpen
  \bibfield  {author} {\bibinfo {author} {\bibfnamefont {K.}~\bibnamefont
  {Lee}}, \bibinfo {author} {\bibfnamefont {{\`E}.~D.}\ \bibnamefont {Murray}},
  \bibinfo {author} {\bibfnamefont {L.}~\bibnamefont {Kong}}, \bibinfo {author}
  {\bibfnamefont {B.~I.}\ \bibnamefont {Lundqvist}}, \ and\ \bibinfo {author}
  {\bibfnamefont {D.~C.}\ \bibnamefont {Langreth}},\ }\href {\doibase
  10.1103/PhysRevB.82.081101} {\bibfield  {journal} {\bibinfo  {journal} {Phys.
  Rev. B}\ }\textbf {\bibinfo {volume} {82}},\ \bibinfo {pages} {081101}
  (\bibinfo {year} {2010}{\natexlab{a}})}\BibitemShut {NoStop}%
\bibitem [{\citenamefont {Berland}\ and\ \citenamefont
  {Hyldgaard}(2014)}]{berland14p035412}%
  \BibitemOpen
  \bibfield  {author} {\bibinfo {author} {\bibfnamefont {K.}~\bibnamefont
  {Berland}}\ and\ \bibinfo {author} {\bibfnamefont {P.}~\bibnamefont
  {Hyldgaard}},\ }\href {\doibase 10.1103/PhysRevB.89.035412} {\bibfield
  {journal} {\bibinfo  {journal} {Phys. Rev. B}\ }\textbf {\bibinfo {volume}
  {89}},\ \bibinfo {pages} {035412} (\bibinfo {year} {2014})}\BibitemShut
  {NoStop}%
\bibitem [{\citenamefont {Berland}\ \emph {et~al.}(2014)\citenamefont
  {Berland}, \citenamefont {Arter}, \citenamefont {Cooper}, \citenamefont
  {Lee}, \citenamefont {Lundqvist}, \citenamefont {Schr{\"o}der}, \citenamefont
  {Thonhauser},\ and\ \citenamefont {Hyldgaard}}]{bearcoleluscthhy14}%
  \BibitemOpen
  \bibfield  {author} {\bibinfo {author} {\bibfnamefont {K.}~\bibnamefont
  {Berland}}, \bibinfo {author} {\bibfnamefont {C.~A.}\ \bibnamefont {Arter}},
  \bibinfo {author} {\bibfnamefont {V.~R.}\ \bibnamefont {Cooper}}, \bibinfo
  {author} {\bibfnamefont {K.}~\bibnamefont {Lee}}, \bibinfo {author}
  {\bibfnamefont {B.~I.}\ \bibnamefont {Lundqvist}}, \bibinfo {author}
  {\bibfnamefont {E.}~\bibnamefont {Schr{\"o}der}}, \bibinfo {author}
  {\bibfnamefont {T.}~\bibnamefont {Thonhauser}}, \ and\ \bibinfo {author}
  {\bibfnamefont {P.}~\bibnamefont {Hyldgaard}},\ }\href@noop {} {\bibfield
  {journal} {\bibinfo  {journal} {J. Chem. Phys.}\ }\textbf {\bibinfo {volume}
  {140}},\ \bibinfo {pages} {18A539} (\bibinfo {year} {2014})}\BibitemShut
  {NoStop}%
\bibitem [{\citenamefont {Berland}\ \emph {et~al.}(2015)\citenamefont
  {Berland}, \citenamefont {Cooper}, \citenamefont {Lee}, \citenamefont
  {Schr{\"o}der}, \citenamefont {Thonhauser}, \citenamefont {Hyldgaard},\ and\
  \citenamefont {Lundqvist}}]{berland15p066501}%
  \BibitemOpen
  \bibfield  {author} {\bibinfo {author} {\bibfnamefont {K.}~\bibnamefont
  {Berland}}, \bibinfo {author} {\bibfnamefont {V.~R.}\ \bibnamefont {Cooper}},
  \bibinfo {author} {\bibfnamefont {K.}~\bibnamefont {Lee}}, \bibinfo {author}
  {\bibfnamefont {E.}~\bibnamefont {Schr{\"o}der}}, \bibinfo {author}
  {\bibfnamefont {T.}~\bibnamefont {Thonhauser}}, \bibinfo {author}
  {\bibfnamefont {P.}~\bibnamefont {Hyldgaard}}, \ and\ \bibinfo {author}
  {\bibfnamefont {B.~I.}\ \bibnamefont {Lundqvist}},\ }\href {\doibase
  10.1088/0034-4885/78/6/066501} {\bibfield  {journal} {\bibinfo  {journal}
  {Rep. Prog. Phys.}\ }\textbf {\bibinfo {volume} {78}},\ \bibinfo {pages}
  {066501} (\bibinfo {year} {2015})}\BibitemShut {NoStop}%
\bibitem [{\citenamefont {Chakarova-K\"{a}ck}\ \emph
  {et~al.}(2006)\citenamefont {Chakarova-K\"{a}ck}, \citenamefont
  {Schr\"{o}der}, \citenamefont {Lundqvist},\ and\ \citenamefont
  {Langreth}}]{chakarova-kack06p146107}%
  \BibitemOpen
  \bibfield  {author} {\bibinfo {author} {\bibfnamefont {S.~D.}\ \bibnamefont
  {Chakarova-K\"{a}ck}}, \bibinfo {author} {\bibfnamefont {E.}~\bibnamefont
  {Schr\"{o}der}}, \bibinfo {author} {\bibfnamefont {B.~I.}\ \bibnamefont
  {Lundqvist}}, \ and\ \bibinfo {author} {\bibfnamefont {D.~C.}\ \bibnamefont
  {Langreth}},\ }\href {\doibase 10.1103/PhysRevLett.96.146107} {\bibfield
  {journal} {\bibinfo  {journal} {Phys. Rev. Lett.}\ }\textbf {\bibinfo
  {volume} {96}},\ \bibinfo {pages} {146107} (\bibinfo {year}
  {2006})}\BibitemShut {NoStop}%
\bibitem [{\citenamefont {Chakarova-K{\"a}ck}\ \emph
  {et~al.}(2006)\citenamefont {Chakarova-K{\"a}ck}, \citenamefont {Borck},
  \citenamefont {Schr\"{o}der},\ and\ \citenamefont
  {Lundqvist}}]{chakarova-kack06p155402}%
  \BibitemOpen
  \bibfield  {author} {\bibinfo {author} {\bibfnamefont {S.~D.}\ \bibnamefont
  {Chakarova-K{\"a}ck}}, \bibinfo {author} {\bibfnamefont {{\O}.}~\bibnamefont
  {Borck}}, \bibinfo {author} {\bibfnamefont {E.}~\bibnamefont {Schr\"{o}der}},
  \ and\ \bibinfo {author} {\bibfnamefont {B.~I.}\ \bibnamefont {Lundqvist}},\
  }\href {\doibase 10.1103/PhysRevB.74.155402} {\bibfield  {journal} {\bibinfo
  {journal} {Phys. Rev. B.}\ }\textbf {\bibinfo {volume} {74}},\ \bibinfo
  {pages} {155402} (\bibinfo {year} {2006})}\BibitemShut {NoStop}%
\bibitem [{\citenamefont {Berland}\ \emph {et~al.}(2011)\citenamefont
  {Berland}, \citenamefont {Chakarova-K{\"a}ck}, \citenamefont {Cooper},
  \citenamefont {Langreth},\ and\ \citenamefont
  {Schr{\"o}der}}]{berland11p135001}%
  \BibitemOpen
  \bibfield  {author} {\bibinfo {author} {\bibfnamefont {K.}~\bibnamefont
  {Berland}}, \bibinfo {author} {\bibfnamefont {S.~D.}\ \bibnamefont
  {Chakarova-K{\"a}ck}}, \bibinfo {author} {\bibfnamefont {V.~R.}\ \bibnamefont
  {Cooper}}, \bibinfo {author} {\bibfnamefont {D.~C.}\ \bibnamefont
  {Langreth}}, \ and\ \bibinfo {author} {\bibfnamefont {E.}~\bibnamefont
  {Schr{\"o}der}},\ }\href {\doibase 10.1088/0953-8984/23/13/135001} {\bibfield
   {journal} {\bibinfo  {journal} {J. Phys.: Condens. Matter}\ }\textbf
  {\bibinfo {volume} {23}},\ \bibinfo {pages} {135001} (\bibinfo {year}
  {2011})}\BibitemShut {NoStop}%
\bibitem [{\citenamefont {Le}\ \emph {et~al.}(2012)\citenamefont {Le},
  \citenamefont {Kara}, \citenamefont {Schr{\"o}der}, \citenamefont
  {Hyldgaard},\ and\ \citenamefont {Rahman}}]{le12p424210}%
  \BibitemOpen
  \bibfield  {author} {\bibinfo {author} {\bibfnamefont {D.}~\bibnamefont
  {Le}}, \bibinfo {author} {\bibfnamefont {A.}~\bibnamefont {Kara}}, \bibinfo
  {author} {\bibfnamefont {E.}~\bibnamefont {Schr{\"o}der}}, \bibinfo {author}
  {\bibfnamefont {P.}~\bibnamefont {Hyldgaard}}, \ and\ \bibinfo {author}
  {\bibfnamefont {T.~S.}\ \bibnamefont {Rahman}},\ }\href {\doibase
  10.1088/0953-8984/24/42/424210} {\bibfield  {journal} {\bibinfo  {journal}
  {J. Phys.: Condens. Matter}\ }\textbf {\bibinfo {volume} {24}},\ \bibinfo
  {pages} {424210} (\bibinfo {year} {2012})}\BibitemShut {NoStop}%
\bibitem [{\citenamefont {{\AA}kesson}\ \emph {et~al.}(2012)\citenamefont
  {{\AA}kesson}, \citenamefont {Sundborg}, \citenamefont {Wahlstr{\"o}m},\ and\
  \citenamefont {Schr{\"o}der}}]{akesson12p174702}%
  \BibitemOpen
  \bibfield  {author} {\bibinfo {author} {\bibfnamefont {J.}~\bibnamefont
  {{\AA}kesson}}, \bibinfo {author} {\bibfnamefont {O.}~\bibnamefont
  {Sundborg}}, \bibinfo {author} {\bibfnamefont {O.}~\bibnamefont
  {Wahlstr{\"o}m}}, \ and\ \bibinfo {author} {\bibfnamefont {E.}~\bibnamefont
  {Schr{\"o}der}},\ }\href {\doibase 10.1063/1.4764356} {\bibfield  {journal}
  {\bibinfo  {journal} {J. Chem. Phys.}\ }\textbf {\bibinfo {volume} {137}},\
  \bibinfo {pages} {174702} (\bibinfo {year} {2012})}\BibitemShut {NoStop}%
\bibitem [{\citenamefont {Schr\"oder}(2013)}]{schroder13p871706}%
  \BibitemOpen
  \bibfield  {author} {\bibinfo {author} {\bibfnamefont {E.}~\bibnamefont
  {Schr\"oder}},\ }\href {\doibase 10.1155/2013/871706} {\bibfield  {journal}
  {\bibinfo  {journal} {J. Nanomater.}\ }\textbf {\bibinfo {volume} {2013}},\
  \bibinfo {pages} {871706} (\bibinfo {year} {2013})}\BibitemShut {NoStop}%
\bibitem [{\citenamefont {Londero}\ \emph {et~al.}(2012)\citenamefont
  {Londero}, \citenamefont {Karlson}, \citenamefont {Landahl}, \citenamefont
  {Ostrovskii}, \citenamefont {Rydberg},\ and\ \citenamefont
  {Schr{\"o}der}}]{londero12p424212}%
  \BibitemOpen
  \bibfield  {author} {\bibinfo {author} {\bibfnamefont {E.}~\bibnamefont
  {Londero}}, \bibinfo {author} {\bibfnamefont {E.~K.}\ \bibnamefont
  {Karlson}}, \bibinfo {author} {\bibfnamefont {M.}~\bibnamefont {Landahl}},
  \bibinfo {author} {\bibfnamefont {D.}~\bibnamefont {Ostrovskii}}, \bibinfo
  {author} {\bibfnamefont {J.~D.}\ \bibnamefont {Rydberg}}, \ and\ \bibinfo
  {author} {\bibfnamefont {E.}~\bibnamefont {Schr{\"o}der}},\ }\href {\doibase
  10.1088/0953-8984/24/42/424212} {\bibfield  {journal} {\bibinfo  {journal}
  {J. Phys.: Condens. Matter}\ }\textbf {\bibinfo {volume} {24}},\ \bibinfo
  {pages} {424212} (\bibinfo {year} {2012})}\BibitemShut {NoStop}%
\bibitem [{\citenamefont {White}\ and\ \citenamefont
  {Proctor}(1997)}]{white97p1239}%
  \BibitemOpen
  \bibfield  {author} {\bibinfo {author} {\bibfnamefont {R.~F.}\ \bibnamefont
  {White}}\ and\ \bibinfo {author} {\bibfnamefont {S.~P.}\ \bibnamefont
  {Proctor}},\ }\href@noop {} {\bibfield  {journal} {\bibinfo  {journal}
  {Lancet}\ }\textbf {\bibinfo {volume} {349}},\ \bibinfo {pages} {1239}
  (\bibinfo {year} {1997})}\BibitemShut {NoStop}%
\bibitem [{GPA(2016)}]{GPAWhttp}%
  \BibitemOpen
  \href@noop {} {\enquote {\bibinfo {title} {Open-source, grid-based
  {PAW}-method {DFT} code {GPAW}},}\ }\bibinfo {howpublished}
  {\url{http://wiki.fysik.dtu.dk/gpaw/}} (\bibinfo {year} {2016})\BibitemShut
  {NoStop}%
\bibitem [{\citenamefont {Enkovaara}\ \emph {et~al.}(2010)\citenamefont
  {Enkovaara}, \citenamefont {Rostgaard}, \citenamefont {Mortensen},
  \citenamefont {Chen}, \citenamefont {Du{\l}ak}, \citenamefont {Ferrighi},
  \citenamefont {Gavnholt}, \citenamefont {Glinsvad}, \citenamefont {Haikola},
  \citenamefont {Hansen}, \citenamefont {Kristoffersen}, \citenamefont
  {Kuisma}, \citenamefont {Larsen}, \citenamefont {Lehtovaara}, \citenamefont
  {Ljungberg}, \citenamefont {Lopez-Acevedo}, \citenamefont {Moses},
  \citenamefont {Ojanen}, \citenamefont {Olsen}, \citenamefont {Petzold},
  \citenamefont {Romero}, \citenamefont {Stausholm-M{\o}ller}, \citenamefont
  {Strange}, \citenamefont {Tritsaris}, \citenamefont {Vanin}, \citenamefont
  {Walter}, \citenamefont {Hammer}, \citenamefont {H\"akkinen}, \citenamefont
  {Madsen}, \citenamefont {Nieminen}, \citenamefont {N{\o}rskov}, \citenamefont
  {Puska}, \citenamefont {Rantala}, \citenamefont {Schi{\o}tz}, \citenamefont
  {Thygesen},\ and\ \citenamefont {Jacobsen}}]{GPAW10}%
  \BibitemOpen
  \bibfield  {author} {\bibinfo {author} {\bibfnamefont {J.}~\bibnamefont
  {Enkovaara}}, \bibinfo {author} {\bibfnamefont {C.}~\bibnamefont
  {Rostgaard}}, \bibinfo {author} {\bibfnamefont {J.~J.}\ \bibnamefont
  {Mortensen}}, \bibinfo {author} {\bibfnamefont {J.}~\bibnamefont {Chen}},
  \bibinfo {author} {\bibfnamefont {M.}~\bibnamefont {Du{\l}ak}}, \bibinfo
  {author} {\bibfnamefont {L.}~\bibnamefont {Ferrighi}}, \bibinfo {author}
  {\bibfnamefont {J.}~\bibnamefont {Gavnholt}}, \bibinfo {author}
  {\bibfnamefont {C.}~\bibnamefont {Glinsvad}}, \bibinfo {author}
  {\bibfnamefont {V.}~\bibnamefont {Haikola}}, \bibinfo {author} {\bibfnamefont
  {H.~A.}\ \bibnamefont {Hansen}}, \bibinfo {author} {\bibfnamefont {H.~H.}\
  \bibnamefont {Kristoffersen}}, \bibinfo {author} {\bibfnamefont
  {M.}~\bibnamefont {Kuisma}}, \bibinfo {author} {\bibfnamefont {A.~H.}\
  \bibnamefont {Larsen}}, \bibinfo {author} {\bibfnamefont {L.}~\bibnamefont
  {Lehtovaara}}, \bibinfo {author} {\bibfnamefont {M.}~\bibnamefont
  {Ljungberg}}, \bibinfo {author} {\bibfnamefont {O.}~\bibnamefont
  {Lopez-Acevedo}}, \bibinfo {author} {\bibfnamefont {P.~G.}\ \bibnamefont
  {Moses}}, \bibinfo {author} {\bibfnamefont {J.}~\bibnamefont {Ojanen}},
  \bibinfo {author} {\bibfnamefont {T.}~\bibnamefont {Olsen}}, \bibinfo
  {author} {\bibfnamefont {V.}~\bibnamefont {Petzold}}, \bibinfo {author}
  {\bibfnamefont {N.~A.}\ \bibnamefont {Romero}}, \bibinfo {author}
  {\bibfnamefont {J.}~\bibnamefont {Stausholm-M{\o}ller}}, \bibinfo {author}
  {\bibfnamefont {M.}~\bibnamefont {Strange}}, \bibinfo {author} {\bibfnamefont
  {G.~A.}\ \bibnamefont {Tritsaris}}, \bibinfo {author} {\bibfnamefont
  {M.}~\bibnamefont {Vanin}}, \bibinfo {author} {\bibfnamefont
  {M.}~\bibnamefont {Walter}}, \bibinfo {author} {\bibfnamefont
  {B.}~\bibnamefont {Hammer}}, \bibinfo {author} {\bibfnamefont
  {H.}~\bibnamefont {H\"akkinen}}, \bibinfo {author} {\bibfnamefont {G.~K.~H.}\
  \bibnamefont {Madsen}}, \bibinfo {author} {\bibfnamefont {R.~M.}\
  \bibnamefont {Nieminen}}, \bibinfo {author} {\bibfnamefont {J.~K.}\
  \bibnamefont {N{\o}rskov}}, \bibinfo {author} {\bibfnamefont
  {M.}~\bibnamefont {Puska}}, \bibinfo {author} {\bibfnamefont {T.~T.}\
  \bibnamefont {Rantala}}, \bibinfo {author} {\bibfnamefont {J.}~\bibnamefont
  {Schi{\o}tz}}, \bibinfo {author} {\bibfnamefont {K.~S.}\ \bibnamefont
  {Thygesen}}, \ and\ \bibinfo {author} {\bibfnamefont {K.~W.}\ \bibnamefont
  {Jacobsen}},\ }\href {\doibase 10.1088/0953-8984/22/25/253202} {\bibfield
  {journal} {\bibinfo  {journal} {J. Phys.: Condens. Matter}\ }\textbf
  {\bibinfo {volume} {22}},\ \bibinfo {pages} {253202} (\bibinfo {year}
  {2010})}\BibitemShut {NoStop}%
\bibitem [{\citenamefont {Bahn}\ and\ \citenamefont {Jacobsen}(2002)}]{ASE02}%
  \BibitemOpen
  \bibfield  {author} {\bibinfo {author} {\bibfnamefont {S.~R.}\ \bibnamefont
  {Bahn}}\ and\ \bibinfo {author} {\bibfnamefont {K.~W.}\ \bibnamefont
  {Jacobsen}},\ }\href {\doibase 10.1109/5992.998641} {\bibfield  {journal}
  {\bibinfo  {journal} {Comput. Sci. Eng.}\ }\textbf {\bibinfo {volume} {4}},\
  \bibinfo {pages} {56} (\bibinfo {year} {2002})}\BibitemShut {NoStop}%
\bibitem [{ASE(2016)}]{ASEhttp}%
  \BibitemOpen
  \href@noop {} {\enquote {\bibinfo {title} {Python-based atomic simulation
  environment {ASE}},}\ }\bibinfo {howpublished}
  {\url{https://wiki.fysik.dtu.dk/ase/}} (\bibinfo {year} {2016})\BibitemShut
  {NoStop}%
\bibitem [{QEh(2016)}]{QEhttp}%
  \BibitemOpen
  \href@noop {} {\enquote {\bibinfo {title} {Open-source {DFT} code {QE}},}\
  }\bibinfo {howpublished} {\url{http://www.quantum-espresso.org/}} (\bibinfo
  {year} {2016})\BibitemShut {NoStop}%
\bibitem [{\citenamefont {Giannozzi}\ \emph {et~al.}(2009)\citenamefont
  {Giannozzi}, \citenamefont {Baroni}, \citenamefont {Bonini}, \citenamefont
  {Calandra}, \citenamefont {Car}, \citenamefont {Cavazzoni}, \citenamefont
  {Ceresoli}, \citenamefont {Chiarotti}, \citenamefont {Cococcioni},
  \citenamefont {Dabo}, \citenamefont {Corso}, \citenamefont {Fabris},
  \citenamefont {Fratesi}, \citenamefont {de~Gironcoli}, \citenamefont
  {Gebauer}, \citenamefont {Gerstmann}, \citenamefont {Gougoussis},
  \citenamefont {Kokalj}, \citenamefont {Lazzeri}, \citenamefont
  {Martin-Samos}, \citenamefont {Marzari}, \citenamefont {Mauri}, \citenamefont
  {Mazzarello}, \citenamefont {Paolini}, \citenamefont {Pasquarello},
  \citenamefont {Paulatto}, \citenamefont {Sbraccia}, \citenamefont {Scandolo},
  \citenamefont {Sclauzero}, \citenamefont {Seitsonen}, \citenamefont
  {Smogunov}, \citenamefont {Umari},\ and\ \citenamefont
  {Wentzcovitch}}]{espresso}%
  \BibitemOpen
  \bibfield  {author} {\bibinfo {author} {\bibfnamefont {P.}~\bibnamefont
  {Giannozzi}}, \bibinfo {author} {\bibfnamefont {S.}~\bibnamefont {Baroni}},
  \bibinfo {author} {\bibfnamefont {N.}~\bibnamefont {Bonini}}, \bibinfo
  {author} {\bibfnamefont {M.}~\bibnamefont {Calandra}}, \bibinfo {author}
  {\bibfnamefont {R.}~\bibnamefont {Car}}, \bibinfo {author} {\bibfnamefont
  {C.}~\bibnamefont {Cavazzoni}}, \bibinfo {author} {\bibfnamefont
  {D.}~\bibnamefont {Ceresoli}}, \bibinfo {author} {\bibfnamefont {G.~L.}\
  \bibnamefont {Chiarotti}}, \bibinfo {author} {\bibfnamefont {M.}~\bibnamefont
  {Cococcioni}}, \bibinfo {author} {\bibfnamefont {I.}~\bibnamefont {Dabo}},
  \bibinfo {author} {\bibfnamefont {A.~D.}\ \bibnamefont {Corso}}, \bibinfo
  {author} {\bibfnamefont {S.}~\bibnamefont {Fabris}}, \bibinfo {author}
  {\bibfnamefont {G.}~\bibnamefont {Fratesi}}, \bibinfo {author} {\bibfnamefont
  {S.}~\bibnamefont {de~Gironcoli}}, \bibinfo {author} {\bibfnamefont
  {R.}~\bibnamefont {Gebauer}}, \bibinfo {author} {\bibfnamefont
  {U.}~\bibnamefont {Gerstmann}}, \bibinfo {author} {\bibfnamefont
  {C.}~\bibnamefont {Gougoussis}}, \bibinfo {author} {\bibfnamefont
  {A.}~\bibnamefont {Kokalj}}, \bibinfo {author} {\bibfnamefont
  {M.}~\bibnamefont {Lazzeri}}, \bibinfo {author} {\bibfnamefont
  {L.}~\bibnamefont {Martin-Samos}}, \bibinfo {author} {\bibfnamefont
  {N.}~\bibnamefont {Marzari}}, \bibinfo {author} {\bibfnamefont
  {F.}~\bibnamefont {Mauri}}, \bibinfo {author} {\bibfnamefont
  {R.}~\bibnamefont {Mazzarello}}, \bibinfo {author} {\bibfnamefont
  {S.}~\bibnamefont {Paolini}}, \bibinfo {author} {\bibfnamefont
  {A.}~\bibnamefont {Pasquarello}}, \bibinfo {author} {\bibfnamefont
  {L.}~\bibnamefont {Paulatto}}, \bibinfo {author} {\bibfnamefont
  {C.}~\bibnamefont {Sbraccia}}, \bibinfo {author} {\bibfnamefont
  {S.}~\bibnamefont {Scandolo}}, \bibinfo {author} {\bibfnamefont
  {G.}~\bibnamefont {Sclauzero}}, \bibinfo {author} {\bibfnamefont {A.~P.}\
  \bibnamefont {Seitsonen}}, \bibinfo {author} {\bibfnamefont {A.}~\bibnamefont
  {Smogunov}}, \bibinfo {author} {\bibfnamefont {P.}~\bibnamefont {Umari}}, \
  and\ \bibinfo {author} {\bibfnamefont {R.~M.}\ \bibnamefont {Wentzcovitch}},\
  }\href {\doibase 10.1088/0953-8984/21/39/395502} {\bibfield  {journal}
  {\bibinfo  {journal} {J. Phys.: Condens. Matter}\ }\textbf {\bibinfo {volume}
  {21}},\ \bibinfo {pages} {395502} (\bibinfo {year} {2009})}\BibitemShut
  {NoStop}%
\bibitem [{\citenamefont {Rom\'{a}n-P\'{e}rez}\ and\ \citenamefont
  {Soler}(2009)}]{roso09}%
  \BibitemOpen
  \bibfield  {author} {\bibinfo {author} {\bibfnamefont {G.}~\bibnamefont
  {Rom\'{a}n-P\'{e}rez}}\ and\ \bibinfo {author} {\bibfnamefont {J.~M.}\
  \bibnamefont {Soler}},\ }\href {\doibase 10.1103/PhysRevLett.103.096102}
  {\bibfield  {journal} {\bibinfo  {journal} {Phys. Rev. Lett.}\ }\textbf
  {\bibinfo {volume} {103}},\ \bibinfo {pages} {096102} (\bibinfo {year}
  {2009})}\BibitemShut {NoStop}%
\bibitem [{\citenamefont {Ericsson}\ \emph {et~al.}(2016)\citenamefont
  {Ericsson}, \citenamefont {Husmark}, \citenamefont {Mathiesen}, \citenamefont
  {Sepahvand}, \citenamefont {Borck}, \citenamefont {Gunnarsson}, \citenamefont
  {Lydmark},\ and\ \citenamefont {Schr\"oder}}]{ericsson16preprint}%
  \BibitemOpen
  \bibfield  {author} {\bibinfo {author} {\bibfnamefont {J.}~\bibnamefont
  {Ericsson}}, \bibinfo {author} {\bibfnamefont {T.}~\bibnamefont {Husmark}},
  \bibinfo {author} {\bibfnamefont {C.}~\bibnamefont {Mathiesen}}, \bibinfo
  {author} {\bibfnamefont {B.}~\bibnamefont {Sepahvand}}, \bibinfo {author}
  {\bibfnamefont {{\O}.}~\bibnamefont {Borck}}, \bibinfo {author}
  {\bibfnamefont {L.}~\bibnamefont {Gunnarsson}}, \bibinfo {author}
  {\bibfnamefont {P.}~\bibnamefont {Lydmark}}, \ and\ \bibinfo {author}
  {\bibfnamefont {E.}~\bibnamefont {Schr\"oder}},\ }\href@noop {} {\enquote
  {\bibinfo {title} {Involving high school students in computational physics
  university research: {T}heory calculations of toluene adsorbed on
  graphene},}\ } (\bibinfo {year} {2016}),\ \bibinfo {note}
  {preprint}\BibitemShut {NoStop}%
\bibitem [{\citenamefont {Monkenbusch}\ and\ \citenamefont
  {Stockmeyer}(1981)}]{monkenbusch81p442}%
  \BibitemOpen
  \bibfield  {author} {\bibinfo {author} {\bibfnamefont {M.}~\bibnamefont
  {Monkenbusch}}\ and\ \bibinfo {author} {\bibfnamefont {R.}~\bibnamefont
  {Stockmeyer}},\ }\href@noop {} {\bibfield  {journal} {\bibinfo  {journal}
  {Ber. Bunsenges. Phys. Chem.}\ }\textbf {\bibinfo {volume} {85}},\ \bibinfo
  {pages} {442} (\bibinfo {year} {1981})}\BibitemShut {NoStop}%
\bibitem [{\citenamefont {Ulbricht}\ \emph {et~al.}(2006)\citenamefont
  {Ulbricht}, \citenamefont {Zacharia}, \citenamefont {Cindir},\ and\
  \citenamefont {Hertel}}]{ulbricht06p2931}%
  \BibitemOpen
  \bibfield  {author} {\bibinfo {author} {\bibfnamefont {H.}~\bibnamefont
  {Ulbricht}}, \bibinfo {author} {\bibfnamefont {R.}~\bibnamefont {Zacharia}},
  \bibinfo {author} {\bibfnamefont {N.}~\bibnamefont {Cindir}}, \ and\ \bibinfo
  {author} {\bibfnamefont {T.}~\bibnamefont {Hertel}},\ }\href@noop {}
  {\bibfield  {journal} {\bibinfo  {journal} {Carbon}\ }\textbf {\bibinfo
  {volume} {44}},\ \bibinfo {pages} {2931} (\bibinfo {year}
  {2006})}\BibitemShut {NoStop}%
\bibitem [{\citenamefont {Zacharia}\ \emph {et~al.}(2004)\citenamefont
  {Zacharia}, \citenamefont {Ulbricht},\ and\ \citenamefont
  {Hertel}}]{zacharia}%
  \BibitemOpen
  \bibfield  {author} {\bibinfo {author} {\bibfnamefont {R.}~\bibnamefont
  {Zacharia}}, \bibinfo {author} {\bibfnamefont {H.}~\bibnamefont {Ulbricht}},
  \ and\ \bibinfo {author} {\bibfnamefont {T.}~\bibnamefont {Hertel}},\ }\href
  {\doibase 10.1103/PhysRevB.69.155406} {\bibfield  {journal} {\bibinfo
  {journal} {Phys. Rev. B}\ }\textbf {\bibinfo {volume} {69}},\ \bibinfo
  {pages} {155406} (\bibinfo {year} {2004})}\BibitemShut {NoStop}%
\bibitem [{\citenamefont {Grimaud}\ \emph {et~al.}(2001)\citenamefont
  {Grimaud}, \citenamefont {Radosavkic}, \citenamefont {Ustaze},\ and\
  \citenamefont {Palmer}}]{grimaud2001}%
  \BibitemOpen
  \bibfield  {author} {\bibinfo {author} {\bibfnamefont {C.-M.}\ \bibnamefont
  {Grimaud}}, \bibinfo {author} {\bibfnamefont {D.}~\bibnamefont {Radosavkic}},
  \bibinfo {author} {\bibfnamefont {S.}~\bibnamefont {Ustaze}}, \ and\ \bibinfo
  {author} {\bibfnamefont {R.}~\bibnamefont {Palmer}},\ }\href {\doibase
  10.1016/S0169-4332(01)00192-1} {\bibfield  {journal} {\bibinfo  {journal}
  {Applied Surface Science}\ }\textbf {\bibinfo {volume} {178}},\ \bibinfo
  {pages} {1} (\bibinfo {year} {2001})}\BibitemShut {NoStop}%
\bibitem [{\citenamefont {Chakarova}\ and\ \citenamefont
  {Schr\"{o}der}(2005)}]{chakarova05p054102}%
  \BibitemOpen
  \bibfield  {author} {\bibinfo {author} {\bibfnamefont {S.~D.}\ \bibnamefont
  {Chakarova}}\ and\ \bibinfo {author} {\bibfnamefont {E.}~\bibnamefont
  {Schr\"{o}der}},\ }\href@noop {} {\bibfield  {journal} {\bibinfo  {journal}
  {J. Chem. Phys.}\ }\textbf {\bibinfo {volume} {122}},\ \bibinfo {pages}
  {054102} (\bibinfo {year} {2005})}\BibitemShut {NoStop}%
\bibitem [{\citenamefont {Chakarova-K{\"a}ck}\ \emph
  {et~al.}(2010)\citenamefont {Chakarova-K{\"a}ck}, \citenamefont {Vojvodic},
  \citenamefont {Kleis}, \citenamefont {Hyldgaard},\ and\ \citenamefont
  {Schr{\"o}der}}]{chakarova-kack10p013017}%
  \BibitemOpen
  \bibfield  {author} {\bibinfo {author} {\bibfnamefont {S.~D.}\ \bibnamefont
  {Chakarova-K{\"a}ck}}, \bibinfo {author} {\bibfnamefont {A.}~\bibnamefont
  {Vojvodic}}, \bibinfo {author} {\bibfnamefont {J.}~\bibnamefont {Kleis}},
  \bibinfo {author} {\bibfnamefont {P.}~\bibnamefont {Hyldgaard}}, \ and\
  \bibinfo {author} {\bibfnamefont {E.}~\bibnamefont {Schr{\"o}der}},\ }\href
  {\doibase 10.1088/1367-2630/12/1/013017} {\bibfield  {journal} {\bibinfo
  {journal} {New J. Phys.}\ }\textbf {\bibinfo {volume} {12}},\ \bibinfo
  {pages} {013017} (\bibinfo {year} {2010})}\BibitemShut {NoStop}%
\bibitem [{\citenamefont {Zhang}\ and\ \citenamefont
  {Yang}(1998)}]{zhang98p890}%
  \BibitemOpen
  \bibfield  {author} {\bibinfo {author} {\bibfnamefont {Y.}~\bibnamefont
  {Zhang}}\ and\ \bibinfo {author} {\bibfnamefont {W.}~\bibnamefont {Yang}},\
  }\href {\doibase 10.1103/PhysRevLett.80.890} {\bibfield  {journal} {\bibinfo
  {journal} {Phys. Rev. Lett.}\ }\textbf {\bibinfo {volume} {80}},\ \bibinfo
  {pages} {890} (\bibinfo {year} {1998})}\BibitemShut {NoStop}%
\bibitem [{\citenamefont {Hyldgaard}\ \emph {et~al.}(2014)\citenamefont
  {Hyldgaard}, \citenamefont {Berland},\ and\ \citenamefont
  {Schr\"oder}}]{hybesc14}%
  \BibitemOpen
  \bibfield  {author} {\bibinfo {author} {\bibfnamefont {P.}~\bibnamefont
  {Hyldgaard}}, \bibinfo {author} {\bibfnamefont {K.}~\bibnamefont {Berland}},
  \ and\ \bibinfo {author} {\bibfnamefont {E.}~\bibnamefont {Schr\"oder}},\
  }\href {\doibase 10.1103/PhysRevB.90.075148} {\bibfield  {journal} {\bibinfo
  {journal} {Phys. Rev. B}\ }\textbf {\bibinfo {volume} {90}},\ \bibinfo
  {pages} {075148} (\bibinfo {year} {2014})}\BibitemShut {NoStop}%
\bibitem [{\citenamefont {Curtiss}\ \emph {et~al.}(1991)\citenamefont
  {Curtiss}, \citenamefont {Raghavachari}, \citenamefont {Trucks},\ and\
  \citenamefont {Pople}}]{curtiss91p7221}%
  \BibitemOpen
  \bibfield  {author} {\bibinfo {author} {\bibfnamefont {L.}~\bibnamefont
  {Curtiss}}, \bibinfo {author} {\bibfnamefont {K.}~\bibnamefont
  {Raghavachari}}, \bibinfo {author} {\bibfnamefont {G.~W.}\ \bibnamefont
  {Trucks}}, \ and\ \bibinfo {author} {\bibfnamefont {J.~A.}\ \bibnamefont
  {Pople}},\ }\href@noop {} {\bibfield  {journal} {\bibinfo  {journal} {J.
  Chem. Phys.}\ }\textbf {\bibinfo {volume} {94}},\ \bibinfo {pages} {7221}
  (\bibinfo {year} {1991})}\BibitemShut {NoStop}%
\bibitem [{\citenamefont {Lee}\ \emph {et~al.}(2010{\natexlab{b}})\citenamefont
  {Lee}, \citenamefont {Morikawa},\ and\ \citenamefont
  {Langreth}}]{lee10p155461}%
  \BibitemOpen
  \bibfield  {author} {\bibinfo {author} {\bibfnamefont {K.}~\bibnamefont
  {Lee}}, \bibinfo {author} {\bibfnamefont {Y.}~\bibnamefont {Morikawa}}, \
  and\ \bibinfo {author} {\bibfnamefont {D.~C.}\ \bibnamefont {Langreth}},\
  }\href {\doibase 10.1103/PhysRevB.82.155461} {\bibfield  {journal} {\bibinfo
  {journal} {Phys. Rev. B}\ }\textbf {\bibinfo {volume} {82}},\ \bibinfo
  {pages} {155461} (\bibinfo {year} {2010}{\natexlab{b}})}\BibitemShut
  {NoStop}%
\bibitem [{\citenamefont {Lee}\ \emph {et~al.}(2011)\citenamefont {Lee},
  \citenamefont {Kelkkanen}, \citenamefont {Berland}, \citenamefont
  {Andersson}, \citenamefont {Langreth}, \citenamefont {Schr{\"o}der},
  \citenamefont {Lundqvist},\ and\ \citenamefont {Hyldgaard}}]{lee11p193408}%
  \BibitemOpen
  \bibfield  {author} {\bibinfo {author} {\bibfnamefont {K.}~\bibnamefont
  {Lee}}, \bibinfo {author} {\bibfnamefont {A.~K.}\ \bibnamefont {Kelkkanen}},
  \bibinfo {author} {\bibfnamefont {K.}~\bibnamefont {Berland}}, \bibinfo
  {author} {\bibfnamefont {S.}~\bibnamefont {Andersson}}, \bibinfo {author}
  {\bibfnamefont {D.~C.}\ \bibnamefont {Langreth}}, \bibinfo {author}
  {\bibfnamefont {E.}~\bibnamefont {Schr{\"o}der}}, \bibinfo {author}
  {\bibfnamefont {B.~I.}\ \bibnamefont {Lundqvist}}, \ and\ \bibinfo {author}
  {\bibfnamefont {P.}~\bibnamefont {Hyldgaard}},\ }\href {\doibase
  10.1103/PhysRevB.84.193408} {\bibfield  {journal} {\bibinfo  {journal} {Phys.
  Rev. B}\ }\textbf {\bibinfo {volume} {84}},\ \bibinfo {pages} {193408}
  (\bibinfo {year} {2011})}\BibitemShut {NoStop}%
\bibitem [{\citenamefont {Chen}\ \emph {et~al.}(2012)\citenamefont {Chen},
  \citenamefont {Al-Saidi},\ and\ \citenamefont {Johnson}}]{chen12p424211}%
  \BibitemOpen
  \bibfield  {author} {\bibinfo {author} {\bibfnamefont {D.~L.}\ \bibnamefont
  {Chen}}, \bibinfo {author} {\bibfnamefont {W.~A.}\ \bibnamefont {Al-Saidi}},
  \ and\ \bibinfo {author} {\bibfnamefont {J.~K.}\ \bibnamefont {Johnson}},\
  }\href {\doibase 10.1088/0953-8984/24/42/424211} {\bibfield  {journal}
  {\bibinfo  {journal} {J Phys: Condens Matter}\ }\textbf {\bibinfo {volume}
  {24}},\ \bibinfo {pages} {424211} (\bibinfo {year} {2012})}\BibitemShut
  {NoStop}%
\bibitem [{\citenamefont {Thonhauser}\ \emph {et~al.}(2015)\citenamefont
  {Thonhauser}, \citenamefont {Zuluaga}, \citenamefont {Arter}, \citenamefont
  {Berland}, \citenamefont {Schr\"oder},\ and\ \citenamefont
  {Hyldgaard}}]{thonhauser15p136402}%
  \BibitemOpen
  \bibfield  {author} {\bibinfo {author} {\bibfnamefont {T.}~\bibnamefont
  {Thonhauser}}, \bibinfo {author} {\bibfnamefont {S.}~\bibnamefont {Zuluaga}},
  \bibinfo {author} {\bibfnamefont {C.~A.}\ \bibnamefont {Arter}}, \bibinfo
  {author} {\bibfnamefont {K.}~\bibnamefont {Berland}}, \bibinfo {author}
  {\bibfnamefont {E.}~\bibnamefont {Schr\"oder}}, \ and\ \bibinfo {author}
  {\bibfnamefont {P.}~\bibnamefont {Hyldgaard}},\ }\href {\doibase
  10.1103/PhysRevLett.115.136402} {\bibfield  {journal} {\bibinfo  {journal}
  {Phys. Rev. Lett.}\ }\textbf {\bibinfo {volume} {115}},\ \bibinfo {pages}
  {136402} (\bibinfo {year} {2015})}\BibitemShut {NoStop}%
\bibitem [{\citenamefont {Klomkliang}\ \emph {et~al.}(2012)\citenamefont
  {Klomkliang}, \citenamefont {Do},\ and\ \citenamefont
  {Nicholson}}]{klomkliang12p5320}%
  \BibitemOpen
  \bibfield  {author} {\bibinfo {author} {\bibfnamefont {N.}~\bibnamefont
  {Klomkliang}}, \bibinfo {author} {\bibfnamefont {D.~D.}\ \bibnamefont {Do}},
  \ and\ \bibinfo {author} {\bibfnamefont {D.}~\bibnamefont {Nicholson}},\
  }\href {\doibase 10.1021/ie300121p} {\bibfield  {journal} {\bibinfo
  {journal} {Industrial and Engineering Chemical Research}\ }\textbf {\bibinfo
  {volume} {51}},\ \bibinfo {pages} {5320} (\bibinfo {year}
  {2012})}\BibitemShut {NoStop}%
\bibitem [{\citenamefont {Kuisma}\ \emph {et~al.}(2016)\citenamefont {Kuisma},
  \citenamefont {Hansson}, \citenamefont {Lindberg}, \citenamefont {Gillberg},
  \citenamefont {Idh},\ and\ \citenamefont {Schr{\"o}der}}]{kuisma16p}%
  \BibitemOpen
  \bibfield  {author} {\bibinfo {author} {\bibfnamefont {E.}~\bibnamefont
  {Kuisma}}, \bibinfo {author} {\bibfnamefont {C.~F.}\ \bibnamefont {Hansson}},
  \bibinfo {author} {\bibfnamefont {T.~B.}\ \bibnamefont {Lindberg}}, \bibinfo
  {author} {\bibfnamefont {C.~A.}\ \bibnamefont {Gillberg}}, \bibinfo {author}
  {\bibfnamefont {S.}~\bibnamefont {Idh}}, \ and\ \bibinfo {author}
  {\bibfnamefont {E.}~\bibnamefont {Schr{\"o}der}},\ }\href@noop {} {\bibfield
  {journal} {\bibinfo  {journal} {J. Chem. Phys.}\ }\textbf {\bibinfo {volume}
  {144}},\ \bibinfo {pages} {184704} (\bibinfo {year} {2016})}\BibitemShut
  {NoStop}%
\end{thebibliography}

%
\end{document}